\documentclass{appolb}
\usepackage{amsmath}
\usepackage{graphicx}
\usepackage{psfrag}

\begin{document}
\title{Ensemble of causal trees
}
\author{Piotr Bialas
\address{Inst. of Physics, Jagellonian University\\
ul. Reymonta 4, 30-059 Krakow}
}
\maketitle
\begin{abstract}
  We discuss the geometry of trees endowed with a causal structure
  using the conventional framework of equilibrium statistical
  mechanics.  We show how this ensemble is related to popular growing
  network models. In particular we demonstrate that on a class of {\em
    afine} attachment kernels the two models are identical but they
  can differ substantially for other choice of weights. We show that
  causal trees exhibit condensation even for asymptotically linear
  kernels.  We derive general formulae describing the degree
  distribution, the ancestor-descendant correlation and the
  probability a randomly chosen node lives at a given geodesic
  distance from the root. It is shown that the Hausdorff dimension
  $d_H$ of the causal networks is generically infinite.
\end{abstract}
  
\section{Introduction}
The study of networks is becoming increasingly popular (for a recent
review see \eg \cite{rev} and also \cite{krzproc}). The main reason
for that is an emergence of great wealth of data on Internet, WWW,
science citation networks, cell metabolism networks and so on.  Most
of those networks (if not all) exhibit features that are not explained
by the classical theory of random graphs due to Erdos and Renyi
\cite{clasic}. Perhaps the most prominent among those features is the
power like degree distribution. Degree of a vertex is the number of
links connected to it. While classical theory predicts a Poissonian
distribution for the degree of a vertex in a random graph, in many of
naturally occurring networks this distribution was found  to have
power--like tails. One way to understand this is based on an ancient
observation ``For unto every one that hath shall be given, and he
shall have abundance: but from him that hath not shall be taken away
even that which he hath.''\cite{mateusz}: a popular web page is more
probable to attract more links to it, a frequently cited paper is more
likely to get more citations and so on. In more modern context this
principle was formulated in \cite{simon} and adopted to the
description of networks in \cite{be} and \cite{ab}.  This is a {\em
  diachronic} approach concentrating on the description of growing
networks.  This is very natural as most of the networks we encounter
are a result of some growth process. The simple models studied in the
literature try to capture the essential features of the growth
mechanism.

One can look at the networks in a different way that is also quite
natural as this is the approach taken in statistical mechanics and
probability theory. In this {\em synchronic} view we treat each
network as a single element of a {\em statistical ensemble}
\cite{bck}. The ensemble is defined by specifying the ``phase space''
that is the class of graphs belonging to it and the weight (or
probability) for each graph in the ensemble.  The probabilities can be
assigned {\em ad hoc} or, what is more interesting, derived from other
principles. In particular it is clear that each growing network model
defines also a statistical ensemble. The ensemble consists of all the
graphs that can be constructed by the specified growth process and the
probability assigned to each graph is the probability of constructing
given graph.  Thus the growth mechanism implicitly defines the
probability for each graph. We find it worthwhile to study what kind
of ensembles can be obtained from the growing network models
\cite{bbjk}. The motivation for this is twofold. First using another
``toolbox'' one can obtain more insight into original models.  Indeed
we are in position to make some general statements about the
correlation functions in growing random networks (GN) models.
Secondly while, as stated, natural networks are usually grown, often
we may not have an access to the growth history and we are effectively
left with the statistical ensemble approach.

In this contribution which is mostly based on \cite{bbjk} we will
describe a statistical ensemble that incorporates the causal structure
inherent in the GN models.  We will show how it relates to the GN
models and derive some results on correlation functions.

\section{Causal trees}
\subsection{Definition}
First we review very quickly the growing random network model
\cite{ab,kr}.  In the simplest version we start with a single vertex
and then at each stage we attach a new vertex to one of the already
existing ones.  The probability for attaching the new $v_{n+1}$ vertex
to some old one $v_i$ depends {\em only} on the degree $n_i$ of the
node $v_i$ and is proportional to $A_{n_i}$, which is called {\em
  attachment kernel}.

It is clear that this process produces a rooted, labeled tree, each
node being labeled by the time at which it was inserted into the
network and the first node being the root. It is also clear that not
all the labelings are possible: the label of the ``father'' must be
smaller then the label of its child. We will call trees that satisfy
this condition {\em causal}.

\begin{figure}
\begin{center}
{\psfrag{t2}[l][l][1]{$1$}
\psfrag{t3}[l][l][1]{$\frac{A_2}{A_1+A_2}$}
\psfrag{t4}[l][l][1]{$\frac{A_1}{2 A_1+A_2}$}
\psfrag{t5}[l][l][1]{$\frac{A_2}{A_3+A_2+2A_1}$}
\noindent\includegraphics[width=11cm]{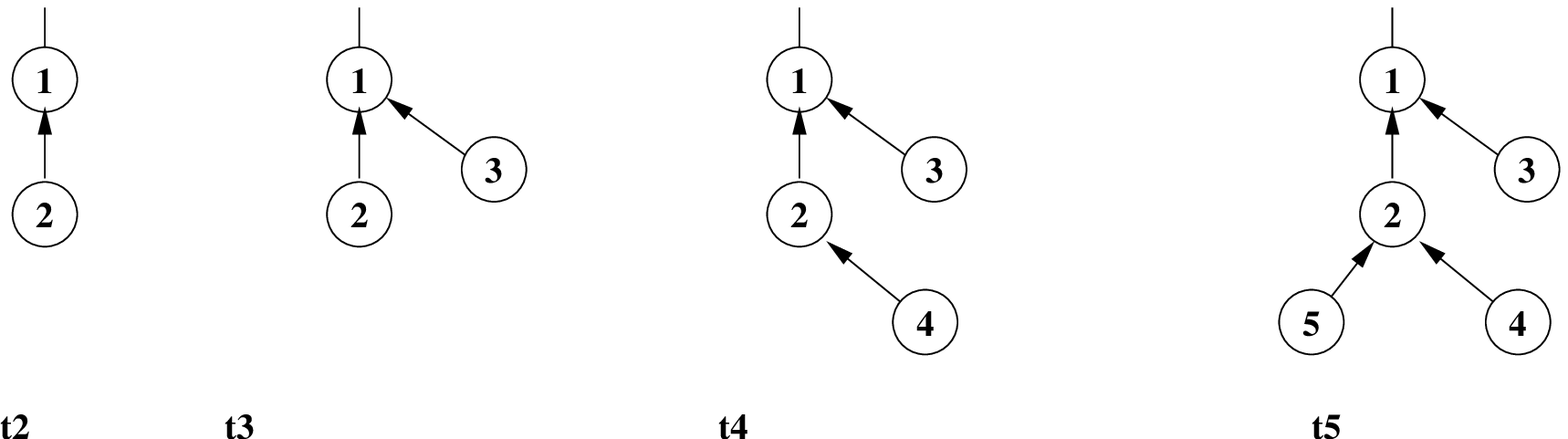}
}\\[10mm]{
\psfrag{t2}[l][l][1]{$1$}
\psfrag{t3}[l][l][1]{$\frac{A_1}{A_1+A_2}$}
\psfrag{t4}[l][l][1]{$\frac{A_2}{A_1+2 A_2}$}
\psfrag{t5}[l][l][1]{$\frac{A_2}{A_3+A_2+2A_1}$}
\includegraphics[width=11cm]{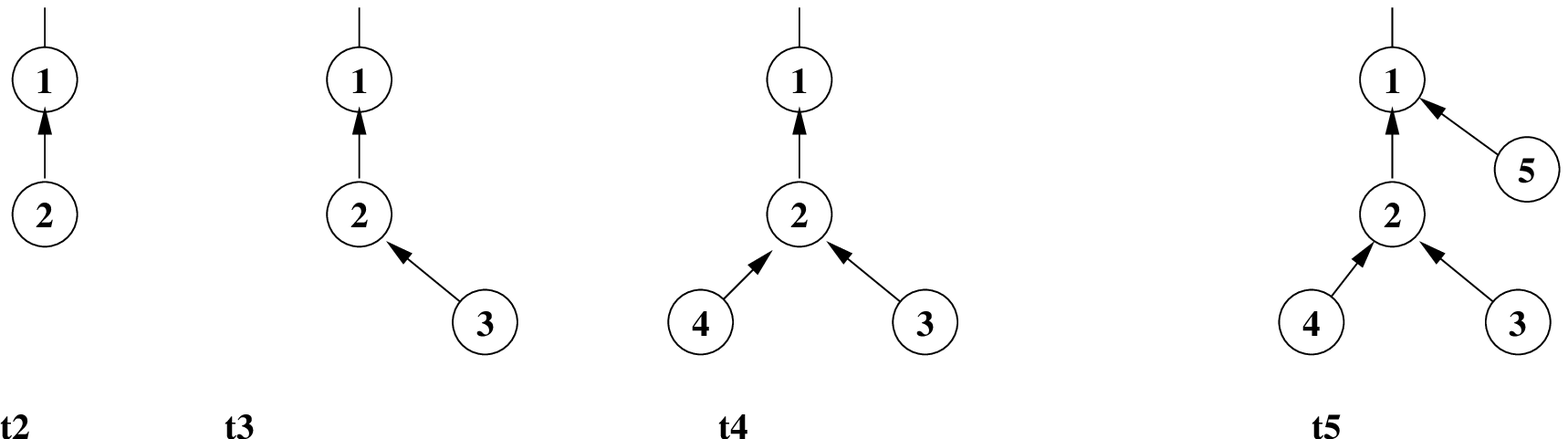}
}
\end{center}
\caption{\label{fig:norm}Two ways of constructing  same (non labeled) tree. The labels below each tree show the probability of obtaining this tree from the precedent one.}
\end{figure}

In order to define the weights we must calculate the probability for
constructing a given tree $T$. To obtain a node $t$ with degree $n$ we
must have attached a new node to the node with degree $n-1$.  The
probability of this happening at the time $t$ is~:
\begin{equation}
P(n|t)=\frac{A_{n-1}}{\sum_{i\in T(t-1)}A_{n_i}}
\end{equation}  
where $T(t-1)$ is the tree at the partial stage of the construction~:
just before attaching the new node $t$.  Unfortunately the normalizing
factor in the denominator in general depends on the exact structure of
the tree $T(t)$. In consequence the overall probability of building a
tree will depend on the way it was constructed, in particular on
the labeling.  This problem is exemplified in the figure~\ref{fig:norm}.
This is situation apart from being impossible to work with is quite
unnatural.  What we would like is to have a weight that depend only
on the nodes degree and do factorise~:
\begin{equation}\label{eq:factor}
\rho(T)=\rho(n_1,\ldots,n_N)=\prod_{i=1}^N q_{n_i}
\end{equation}
It turns out that there exists a class of GN models that is compatible 
with the above requirement. Those are the models with {\em afine}
attachment kernels \ie of the form~:
\begin{equation}\label{eq:afine}
A_{n}=n+\omega,\qquad \omega>-1
\end{equation}
where $\omega$ is a constant. For such kernels the normalization factor 
depends only on the size of the tree~:
\begin{equation}
\sum_i A_{n_i}=\sum_i n_i +\sum_i \omega =2N-2 +N \omega
\end{equation}
For this class of attachment kernels the choice 
\begin{equation}\label{eq:qn}
q_n = q_1\prod_{k=1}^{n-1}A_k
\end{equation}
leads to a model {\em identical} with the original GN model.  For
other kernels we will still define our model by the formulas
\eqref{eq:factor} and \eqref{eq:qn}. In this situation we can only
expect some form of asymptotic or qualitative agreement between the
models if any. In fact we as we will show later the two models can
differ significantly even for very simple non-afine kernels. 

\subsection{Recursion relation}

Most of the properties of the ensemble can be derived from the
canonical partition function~:
\begin{equation}
z_N=\sum_{T}\frac{L(T)}{N!}\rho(T)
\end{equation}
where we sum over all the {\em non labeled } trees and $L(T)$ is the
number of distinct causal labelings of the given tree $T$.

\begin{figure}
\begin{center} 
\psfrag{t1}[c][c]{$T_1$}
\psfrag{t2}[c][c]{$T_2$}
\psfrag{t3}[c][c]{$T_3$}
\includegraphics[width=11cm]{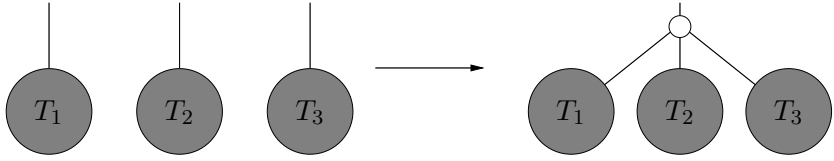}
\end{center}
\caption{\label{fig:lt} Operation $T_1,\ldots,T_k \rightarrow T_1\otimes\cdots\otimes T_k$}
\end{figure}

We start by deriving a recursion relation for $L(T)$.  It is convenient
at this stage to change to the {\em planted} ensemble. This amount to
attaching to the root vertex an additional link~: {\em a stem}. This
does not change in any way the properties of the
ensemble in the large $N$ limit but makes the calculations easier.
From $k$ planted trees $T_1,\ldots,T_k$ we can construct a new tree 
$T=T_1\otimes\cdots\otimes T_k$ by  joining together the stems at a new node (see figure~\ref{fig:lt}).
The number of causal labelings of the resulting tree is~:
\begin{equation}\label{eq:Lrec}
L(T_1\oplus\cdots\oplus T_k)=\frac{N!}{N_1!\cdots N_k!}\frac{1}{k!}
L(T_1)\cdots L(T_k)
\end{equation}
with 
$N_1+\cdots+N_k=N$. 
One has to give $N+1$ labels to the nodes of the 
compound tree. However, the smallest label must be attached to the root. 
The remaining $N$ labels are arbitrarily distributed among the trees. 
This is the origin of the the multinomial factor. Permuting the trees 
$T_i$ does not change the compound tree. This explains 
the presence of the factor $1/k!$. 

Because of the property \eqref{eq:factor} the weights
of the new tree obviously factorise as~:
\begin{equation}\label{eq:rhorec}
\rho(T_1\oplus\cdots\oplus T_k)=p_{k+1}\rho(T_1)\cdots\rho(T_k)
\end{equation}

The partition function $Z_{N+1}$ can be constructed by summing the trees 
of size smaller or equal to $N$ (see figure~\ref{fig:recr})~:
\begin{figure} 
\begin{center} 
\psfrag{n}[c][c][.7]{$N+1$}
\psfrag{n11}[c][c][.7]{$N$}
\psfrag{n1}[c][c][.7]{$N_1$}
\psfrag{n2}[c][c][.7]{$N_2$}
\psfrag{t1}[c][c][.7]{$N_1$}
\psfrag{t2}[c][c][.7]{$N_2$}
\psfrag{t3}[c][c][.7]{$N_3$}
\includegraphics[width=11cm]{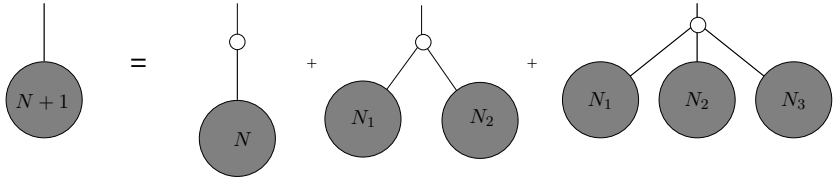}
\end{center}
\caption{\label{fig:recr}Recursion relation for $z_{N+1}$}
\end{figure}
\begin{multline}
  z_{N+1}=\frac{1}{(N+1)!}\sum_{k=1}^\infty
  \sum_{T_1,\ldots,T_k}\!\!\delta_{N_1+\cdots+N_k,N}\times\\
  \times\rho(T_1\oplus\cdots\oplus
  T_k)L(T_1\oplus\cdots\oplus T_k)
\end{multline}
Inserting \eqref{eq:Lrec} and \eqref{eq:rhorec} and rearranging the
terms we arrive at~:
\begin{align}\label{eq:rec}
Z_{N+1}=\frac{1}{N+1}\sum_{k=1}^\infty \frac{p_{k+1}}{k!}
\sum_{N_1,\ldots,N_k}\delta_{N_1+\cdots+ N_k,N}\prod_{i=1}^{k}z_{N_k}  
\end{align}

Adding $z_1=q_1$ and summing both sides of equation \eqref{eq:rec} we get
\begin{equation}        
\sum_N N z_N e^{-N \mu}=
e^{-\mu}\left(\sum_{k=0}^{\infty} \frac{q_{k+1}}{k!}Z(\mu)^k\right)             \end{equation}
where 
\begin{equation}
Z(\mu)=\sum_Nz_Ne^{-\mu N}
\end{equation}
is the grand--canonical partition function. Finally 
\begin{equation}\label{eq:diff}
Z'(\mu) = - \, e^{-\mu} F(Z)
\end{equation}
where 
\begin{equation}\label{eq:defF}
F(Z) = \sum_{k=1}^{\infty} \frac{q_k}{(k-1)!} Z^{k-1}
\end{equation}
Equation \eqref{eq:diff} can be integrated to give
\begin{equation}\label{eq:Z}    
e^{-\mu(Z)} = G(Z) \equiv \int_0^Z\frac{\text{d}x}{F(x)}
\end{equation}
The function $G(Z)$ is a positive monotonically growing function of
Z, bounded from above (one can ignore the trivial case where all 
$q_n$ except $q_1$ and $q_2$ are zero). Hence $\mu$ is bounded
from below: $Z(\mu)$ has a singularity at some $\mu=\bar{\mu}$.
Denote by $\bar{x}$ the radius of convergence of the series $F(Z)$.
The critical value of $\mu$ is given by
\begin{equation}\label{eq:muc}
\bar{\mu} = -\log G(\bar{x})
\end{equation}
This formula holds also when the radius of convergence $\bar{x}$ is
infinite, since all terms in the series \eqref{eq:defF} are positive
and the integral in \eqref{eq:muc} is convergent in all cases of
interest~: $G(\infty)<\infty$. Please note that $\bar\mu$ is a ``free
energy'' density~:
\begin{equation}
\bar\mu=\lim_{N\rightarrow\infty}\frac{1}{N}\log z_N
\end{equation}

\subsection{Degree distribution }

The vertex degree distribution is calculated using~: 
\begin{equation}
\pi_n=q_n\frac{\partial \bar\mu}{\partial q_n}
\end{equation}
which gives
\begin{equation}\label{eq:vdd}
\pi_n = \frac{1}{G(\bar{x})}\frac{q_n}{(n-1)!}
\int_0^{\bar{x}}\kern-2mm\frac{\text{d}x}{F(x)^2}x^{n-1}                
\end{equation}
Again, this formula is also valid when $\bar{x}=\infty$. 
\par
Summing over $n$ and using the definitions of $F$ and $G$ one easily
checks that $\pi_n$ is normalized to unity, as it should. One further 
finds
\begin{equation}\label{eq:sumto2}
\sum_n n \; \pi_n = 2 - \underbrace{\frac{\bar{x}}{F(\bar{x}) G(\bar{x})}           }_c
\end{equation}
On a tree, the r.h.s. should equal 2. This is the case when $F(x)$
diverges at $x=\bar{x}$ and hence $c=0$.  Otherwise one encounters a
pathology (anomaly), which looks similar to that appearing in some
maximally random tree models (and in the so-called balls-in-boxes
model, see \cite{bck,bbj}), where in the large $N$ limit one misses
singular node(s) contributing term(s) of the type
\begin{equation}\label{eq:peakpos}
N^{-1} \delta (n- cN)
\end{equation}
Such nonuniformly behaving terms disappear if one first takes the
$N\rightarrow\infty$ limit in \eqref{eq:sumto2}.  It will be shown
later that the average distance between nodes is finite when
$F(\bar{x}) < \infty$ . This means that singular node(s) - with
unbounded connectivity - are indeed expected to show up. 

\subsection{Condensation}

\begin{figure}
\begin{center}
\psfrag{x}[c][c]{$n$}
\psfrag{y}[c][c][1][-90]{$\pi_n$}
\includegraphics[width=12cm]{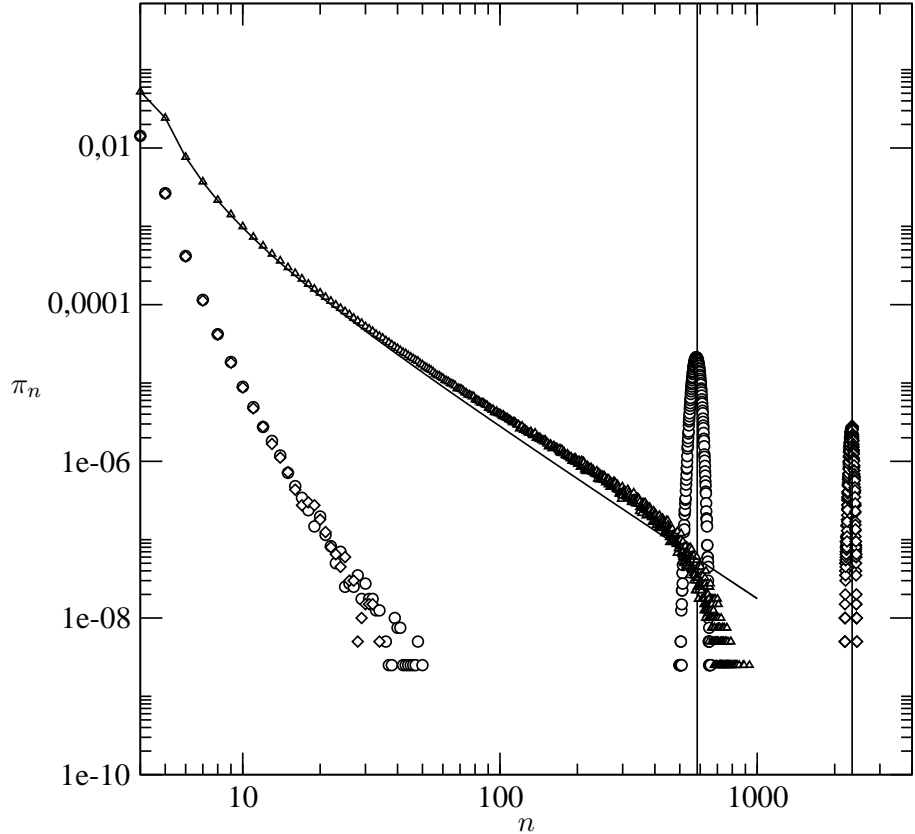}
\end{center}
\caption{\label{fig:cond}Vertex degree distribution for causal trees ( 
  1000 nodes (circles), 4000 nodes (diamonds)) and GN model (4000 nodes
  (triangles))}
\end{figure}

In order to check if the described anomaly really signals an
appearance of a singular vertex we have studied causal trees with
the weights 
\begin{equation}
q_n=\begin{cases}
1 & n\le d\\
(n-d)! & n>d
\end{cases}
\end{equation}
where $d$ is an integer greater or equal to two,
 derived using \eqref{eq:qn} from the {\em delayed} linear attachment kernel
\begin{equation}\label{eq:delayed}
A_n=\begin{cases}
1 & n\le d\\
n-d+1 & n>d
\end{cases}
\end{equation}
In this case the coefficients of the power series \eqref{eq:defF} behave like~:
\begin{equation} 
\frac{(k-d)!}{(k-1)!}\sim \frac{1}{k^{d-1}}\quad\text{for } k\rightarrow\infty 
\end{equation}
The term $c$ in \eqref{eq:vdd} is obviously not zero and its value is
found to be $c\approx 0.584692$.  In the figure~\ref{fig:cond} we have
ploted the result of simulations of the model of causal trees with
1000 and 4000 nodes (circles and diamonds). One can clearly see peaks
corresponding to the singular node. The vertical lines mark the
positions of those peaks predicted from \eqref{eq:peakpos}.  The
perfect agreement confirms our statement that the ``link deficiency''
in \eqref{eq:sumto2} signals the existence of a singular node.

It is interesting to ask what is the shape of the
vertex degree distribution for the growing random model which has
exactly the same kernel \eqref{eq:delayed}.  Of course in this case
the two models are not identical because the kernel is not afine.  The
resulting degree distribution for this model is ploted in the
figure~\ref{fig:cond} (triangles) \cite{akoles}. The continuous line
is the approximate solution taken from \cite{kr}.  The observed
discrepancy is due to finite size effects.  As we can see the
distributions in causal trees and growing networks differ greatly, in
particular there is no singular vertex in the GN model for this choice
of kernel.

\subsection{Degree correlations}

Now, we turn to the calculation of the ancestor-descendant degree 
correlation. It is obvious that an ancestor plays the role of the root
of the subgraph involving all its descendants. One can read from
\eqref{eq:rec} the degree distribution of the root:
\begin{equation}
z_l(N) = \frac{1}{N}\frac{q_l}{(l-1)!} 
\kern-1mm\sum_{N_1,\dots,N_{l-1}}\kern-3mm \delta_{N_1+\cdots+N_{l-1}, N-1} 
\prod_1^{l-1} Z_{N_i}
\end{equation}
Going over to the grand-canonical ensemble one finds:
\begin{equation}
\frac{\text{d}Z_l(\mu)}{\text{d}\mu} = 
-e^{-\mu} \frac{q_l}{(l-1)!} Z^{l-1}(\mu)
\end{equation}
which, taking \eqref{eq:diff} into account and after integration yields
\begin{equation}
Z_l\bigl(\mu(Z)\bigr) =  \frac{q_l}{(l-1)!} 
\int_0^Z \text{d}x \frac{x^{l-1}}{F(x)}
\end{equation}
Using similar arguments one writes the weight of graphs where the 
root has the degree $l$ and its daughter the degree $k$ as
\begin{equation}\begin{split}
z_{kl}(N) = &\frac{q_l}{N (l-2)!}\sum_{N_1,\dots,N_{l-1}} 
\kern-3mm\delta_{N_1+\dots+N_{l-1}, N-1} \times\\
&\times \prod_{i=1}^{l-2} Z_{N_i} z_k(N_{l-1})
\end{split}
\end{equation}
Hence
\begin{equation}
\frac{\text{d}Z_{kl}(\mu)}{\text{d}\mu} = -e^{-\mu} 
\frac{q_l}{(l-2)!} Z^{l-2}(\mu)Z_k(\mu)
\end{equation}
Integrating the above equation one finally obtains 
\begin{multline}\label{eq:2ptcor}
Z_{kl}\bigl(\mu(Z)\bigr) = \frac{q_l}{(l-2)!} \frac{q_k}{(k-1)!}\int_0^Z \text{d}x_2 \frac{x_2^{l-2}}{F(x_2)} 
\int_0^{x_2} \text{d}x_1 \frac{x_1^{k-1}}{F(x_1)}
\end{multline}
which is the conditional probability, up to normalization, that a 
descendant has the degree $k$ when the ancestor's degree is $l$.
The normalization is determined summing over $k$ on the 
r.h.s. above, with the result $(l-1)Z_l(\mu)$. 

\subsection{Fractal dimension}
Repeating over and over  the iteration process 
leading to eq. \eqref{eq:2ptcor} one gets
\begin{multline}\label{eq:rptcor}
Z_{k_1k_2 \dots k_r}(\mu(Z)) =   \\
 \prod_{j=2}^r \frac{q_{k_j}}{(k_j-2)!}\; \frac{q_{k_1}}{(k_1-1)!}
\int_0^Z \text{d}x_r \frac{x_r^{k_r-2}}{F(x_r)} \times  \\
\times \int_0^{x_{r}}\kern-3mm\text{d}x_{r-1} 
\frac{x_{r-1}^{k_{r-1}-2}}{F(x_{r-1})} 
\cdots \int_0^{x_2} \text{d}x_1 \frac{x_1^{k_1-1}}{F(x_1)}\; \; \;
\end{multline}
Summing over node degrees $k_1, k_2, \cdots , k_r$ one obtains 
the weight of all graphs with a point separated by $r$ links from 
the root, \ie the two-point correlation
function $C(r, \mu)$ introduced in sect. 1.2~:
\begin{multline}
C\bigl(r, \mu(Z)\bigr) = \\\int_0^Z \text{d}x_r \frac{F'(x_r)}{F(x_r)} 
\int_0^{x_r} \text{d}x_{r-1} \frac{F'(x_{r-1})}{F(x_{r-1})}\cdots \times\\
\times \int_0^{x_3} \text{d}x_2 \frac{F'(x_2)}{F(x_2)}
\int_0^{x_2} \text{d}x_1 \; \; \; \;
\label{cr} 
\end{multline}
For finite $\bar{x}$, replacing the upper limit 
of integration over $x_1$ by $\bar{x}$ and
performing all integrations, one gets
\begin{equation}
C\bigl(r, \mu(Z)\bigr) \leq \bar{x} \frac{\bigl(\ln F(Z)\bigr)^{r-1}}{(r-1)!}
\end{equation}
Hence, the tail of $C(r, \mu)$ falls at least as fast as a Poissonian.
Consequently $\langle r \rangle_\mu$ grows at most like $\ln{F(Z)}$.
Assuming that $F(z)$ has at most a power singularity 
at $z=\bar{x}$ one concludes that  
\begin{equation}
\langle r \rangle_\mu  \leq 
\mbox{\rm const} \ln{\frac{1}{\delta\mu}}
\end{equation}
and therefore
\begin{equation}
\langle r \rangle_N \leq \mbox{\rm const} \ln{N}
\end{equation}
since $\delta\mu$ scales like $N^{-1}$. The argument is rather
heuristic, but suggestive (see also the examples in the section IIF of
ref~\cite{bbjk}).  It appears that generically the causal trees have
the {\em small-world} property $d_H=\infty$, contrary to the maximum
entropy trees whose generic fractal dimension is finite
\cite{bck,adj,jk}.  This phenomenon is easy to understand
intuitively~: the causal structure suppresses long branches. This can
be seen by noting that along a branch from the root to the leaf no
label permutations are possible, hence a tree with a few long branches
admits much less causal labelings then a ``short fat'' one.

\section{Summary}

We have studied a statistical ensemble of tree graphs endowed with a
causal structure. We have derived some general formulas describing the
degree distribution, the ancestor--descendant correlation, and the
probability that a node lives at a given geodesic distance from the
root.  Using these last results, we have shown that our causal
networks have generically the small--world property \ie their
Hausdorff dimension is infinite. 

We have shown that our model coincides with the growing random model
for a  afine class of attachment kernels. Outside this class however the
models can wildly differ. In particular we have demonstrated that
while condensation of links can be observed in causal trees it is not
to be seen in their growing network analogue (\ie for the same
weights).

\section{Acknowledgments}

This work was supported by Polish State Committee for Scientific
Research (KBN), grant 2P03B 09622 (2002-2004).

\end{document}